\documentclass[11pt,a4paper,reprint,notitlepage,superscriptaddress,aps,prl]{revtex4-1}

\usepackage{amsmath,amssymb,amsfonts}

\usepackage{physics}

\usepackage[svgnames,dvipsnames]{xcolor}
\usepackage{graphicx}
\definecolor{NewBlue}{rgb}{0.1, 0.1, 0.7}
\definecolor{NewRed}{rgb}{0.7, 0.1, 0.1}
\usepackage[colorlinks,
    linkcolor=NewRed,
    citecolor=NewBlue,
    urlcolor=NewRed]{hyperref}

\newcommand{\lb}{\left(}
\newcommand{\rb}{\right)}

\newcommand{\bldk}{\mathbf{k}}

\usepackage[capitalise]{cleveref}

\renewcommand{\t}[1]{\mathrm{{#1}}}
\renewcommand{\phi}{\varphi}

\begin{document}

\title{Acceleration-induced effects in stimulated light-matter interactions}

\author{Barbara \v{S}oda}
\thanks{bsoda@uwaterloo.ca}
\affiliation{Department of Physics, University of Waterloo, Waterloo, ON N2L 3G1, Canada}
\affiliation{Perimeter Institute for Theoretical Physics,
Waterloo, ON N2L 2Y5, Canada}

\author{Vivishek Sudhir}
\affiliation{Department of Mechanical Engineering, Massachusetts Institute of Technology, Cambridge, MA 02139, USA}
\affiliation{LIGO Laboratory, Massachusetts Institute of Technology, Cambridge, MA 02139, USA}

\author{Achim Kempf}
\affiliation{Department of Applied Mathematics, University of Waterloo, Waterloo, ON N2L 3G1, Canada}
\affiliation{Institute for Quantum Computing, University of Waterloo, Waterloo, ON N2L 3G1, Canada}
\affiliation{Perimeter Institute for Theoretical Physics,
Waterloo, ON N2L 2Y5, Canada}
\affiliation{Department of Physics, University of Waterloo, Waterloo, ON N2L 3G1, Canada}

\begin{abstract}
The interaction between light and an atom
proceeds via three paradigmatic mechanisms: 
spontaneous emission, stimulated emission, and
absorption. All three are resonant processes in the sense that they require that the radiation field be 
resonant with the atomic transition.
The non-resonant counterparts of these effects, while necessary to maintain locality of the interaction in principle, 
are usually negligible because their effects tend to average out over multiple cycles of the radiation field. 
This state of affairs does not hold if the atom is accelerated. 
We show that, when accelerated, the non-resonant effects can be made to dominate over the conventional 
resonant effects. 
In fact we show that the non-resonant effects can be vastly enhanced by stimulation, and 
that suitably chosen acceleration can entirely suppress the resonant effects. 
In the class of effects that we study, the Unruh effect is the special case of vanishing stimulation. 
\end{abstract}

\maketitle


\noindent \emph{Introduction. }
Ever since Stern's invention of the molecular beam technique \cite{Stern26} --- which beget contemporary 
atomic physics --- the study of the light-matter interaction has largely been a quest to engineer and increase 
the strength with which individual photons interact with individual atom-like systems. 
Landmark advances over the past century, starting with the observation that two atomic levels can be driven 
by a resonant (classical) electromagnetic field \cite{Rabi37,Rams90}, 
that atoms can be stimulated and optically pumped into desired states \cite{Kast57}, 
that even the electromagnetic vacuum can modify the atomic response \cite{Klepp81,DalCoh82,Milonni}, 
all rely on enhancing the light-matter interaction by \emph{resonating} light with an atomic transition. 
Such resonant effects are described by an interaction of the form \cite{Dirac27,JayCum63},
$a^\dagger \sigma_- + a \sigma_+$, where $\sigma_{\pm}$ raises/lowers the atomic level, and $a^\dagger,a$ 
creates/annihilates a photon. The taxonomy of this so-called Jaynes-Cummings interaction, in its various regimes
of physical interest, has been a driver of quantum optics and atomic physics \cite{AllenEber87,Cohen04,HarRaim06}. 

In principle, the Jaynes-Cummings interaction --- even for a single mode of the electromagnetic field interacting with 
a single two-level atom --- cannot be a complete description. Expressed in terms of the local quantum fields 
that mediate the interaction, the Jaynes-Cummings interaction is non-local. Indeed in quantum optics, the 
Jaynes-Cummings interaction results from an approximation: the exact local interaction between the dipole moment
$\mu$ and field $\phi$, of the form \cite{Rabi37,Dicke54} $\mu (x) \phi(x) \propto (a+a^\dagger) (\sigma_+ + \sigma_-)$, 
is subjected to the rotating wave approximation. 
The rationale being that in a frame rotating at the resonance frequency, terms
such as $a^\dagger \sigma_+$ and $a \sigma_-$ pick up oscillating phase factors that average out in any observation
window that is long compared to an optical cycle. However, at least two instances are known where this
rationale can break down. One is the regime of ultra-strong coupling where these so-called counter-rotating wave 
terms dress the atomic levels so thickly that their effect is significant \cite{ultrastrong19}. 
The other is when the atom is on a trajectory such that the electromagnetic field it sees is no longer monochromatic 
--- due to a chirped Doppler effect --- so that the phase factors may not average out.
A simple trajectory where this happens is when the atom is uniformly 
accelerating \cite{Unruh76,UnWal84,Hu04,Scully18}. 
Indeed, in this context, when the field
is assumed to be in the vacuum state, the accelerating atom is predicted to perceive an apparently thermal 
field --- the Unruh effect \cite{Unruh76,UnWal84,Tak86,CrispMats08,Ben19}. 
The Unruh effect is too weak to be measurable at present.
Nevertheless, it is important because of its correspondence with the Hawking effect \cite{Hawk75} 
(via the equivalence principle); this correspondence relates the behaviour of accelerating atoms 
to quantum effects in gravitational fields.
So despite acceleration being a classical concept, its effect on quantum 
systems can be profound and insightful \cite{Green79,JaeRey97}.

These arguments call for a thorough and general investigation of the full extent to which acceleration can impact the light-matter interaction. 

In this Letter, we present new phenomena at the nexus between non-inertial motion and the light-matter interaction.
We begin by studying 
the effect of stimulation on emission and absorption for an atom in non-inertial motion, generalizing both the spontaneous version of 
this phenomenon (the Unruh effect), as well as the conventional stimulated emission and absorption
of an inertial atom \cite{Dirac27}.
We then show that acceleration impacts not only the counter-rotating terms but also the rotating wave terms. 
In particular, we find that, for certain accelerating trajectories, effects that are due to the usual rotating wave terms can be arbitrarily well suppressed, leading, for example, to ``acceleration-induced transparency". 
As an important practical consequence, the counter-rotating terms can be made to dominate the 
response of an accelerating atom.
These results hint at ways of vastly enhancing quantum effects in non-inertial motion (and possible implications, via the equivalence principle, also for gravity), and may pave a path towards their demonstration in experiments.
They also promise a new route to the physics of counter-rotating-wave terms in the light matter interaction
without ultra-strong coupling.

\emph{Model of light-matter interaction. }
We consider a two-level atom with states $\{\ket{g},\ket{e}\}$ separated by an energy gap $\Omega$ (in units of
$\hbar =1$) in an inertial frame, moving through spacetime that is pervaded by a massless scalar quantum 
field $\hat{\phi}(\vb{x},t)$. The free Hamiltonian for the system is, 
\begin{align*}
	\hat{H}_0 = \Omega \hat{\sigma}_z + \tfrac{1}{2}: (\partial_\mu \hat{\phi})^2 :
		= \Omega \hat{\sigma}_z + \int (\dd k)\, 
			\omega_{\bldk}\, \hat{a}_{\vb{k}}^\dagger \hat{a}_{\vb{k}},
\end{align*}
where, $\hat{\sigma}_z = \tfrac{1}{2}(\dyad{e}{e}- \dyad{g}{g})$,
$:\cdots :$ represents normal order, $\omega_\bldk =  c k$ is the frequency of the field mode whose
annihilation/creation operators are $\hat{a}_{\vb{k}},\hat{a}_{\vb{k}}^\dagger$,
and $(\dd k) = \dd^3 k/[(2\pi)^{3/2}\omega_{\vb{k}}^{1/2}]$ is the Lorentz
invariant integration measure.
The interaction between the two systems, modelled after the minimal coupling
between a charge distribution and the electromagnetic field, is taken to be \cite{UnWal84},
\begin{equation}
	\hat{H}_\t{int} = G\, \hat{\mu}\, \hat{\phi}\left( \vb{x}(\tau),t(\tau) \right),
\end{equation}
where $G$ is the coupling strength, $\hat{\mu}=\dyad{e}{g} + \dyad{g}{e}$, and $\hat{\phi}[\vb{x}(\tau), t(\tau)]$ 
is the field along the detector's trajectory $(\vb{x}(\tau),t(\tau))$ in terms of its proper time $\tau$.

We are, in general, interested in the probability that an initial state $\ket{\psi_\t{i}}$
produces a state $\ket{\psi_\t{f}}$ via the interaction. Moving to the interaction picture
defined with respect to the free Hamiltonian, the interaction Hamiltonian becomes time-dependent,
$\hat{H}_\t{int}(\tau) = G\, \hat{\mu}(\tau) \hat{\phi}(\vb{x}(\tau),t(\tau))$, where,
\begin{align*}
	\mu(\tau) &= e^{i \Omega \tau} \hat{\sigma}_+ + h.c.  \\
	\hat{\phi}(\vb{x},t) &= \int (\dd k)\, 
		\left( e^{-i \omega_{\vb{k}} t +i \vb{k}\cdot \vb{x}} \hat{a}_{\vb{k}} + h.c. \right).
\end{align*}
Here, $\hat{\sigma}_+ = \dyad{e}{g}$ is the raising operator for the atom, and its conjugate, $\hat{\sigma}_- = 
\hat{\sigma}_+^\dagger$ is the lowering operator.
Restricting attention to the weak-coupling regime
the transition amplitude for the process $\ket{\psi_\t{i}} \rightarrow \ket{\psi_\t{f}}$ is (up to a phase),
\begin{align*}
	\mathcal{A}_{\t{i} \rightarrow \t{f}} 
		= \int \mel{\psi_\t{f}}{\hat{H}_\t{int}(\tau)}{\psi_\t{i}} \dd \tau
		= \int (\dd k)\, 
		\mathcal{A}_{\t{i}\rightarrow \t{f}}(\vb{k}),
\end{align*}
where, 
\begin{equation}
\begin{split}
	\mathcal{A}_{\t{i}\rightarrow \t{f}}(\vb{k}) = G \int \dd \tau \bra{\psi_\t{f}}
		&\left( e^{i\Omega \tau}\hat{\sigma}_+ + h.c.\right)\\
		\times &\left( e^{-i \omega_{\vb{k}} t +i \vb{k}\cdot \vb{x}} \hat{a}_{\vb{k}} + h.c. \right) \ket{\psi_\t{i}},
\end{split}
\end{equation}
is the amplitude for the transition $\ket{\psi_i}\rightarrow \ket{\psi_\t{f}}$ mediated by a field quantum of
momentum $\vb{k}$ in proper time $\tau$. 
Defining the integral,
\begin{equation}
	I_\pm(\Omega,\bldk) = \int \dd \tau\, e^{i\Omega \tau \pm i k^\mu x_\mu (\tau)},
\end{equation}
the amplitude can be expressed as,
\begin{align*}
	\mathcal{A}_{\t{i}\rightarrow \t{f}}(\vb{k}) &= 
	G \left[ I_-(\Omega,\bldk) \bra{\psi_\t{f}} \hat{\sigma}_- \hat{a}_{\bldk}^\dagger \ket{\psi_\t{i}} + h.c. \right] \\
	&\quad + G \left[ I_+(\Omega,\bldk) \bra{\psi_\t{f}} \hat{\sigma}_+ \hat{a}_{\bldk}^\dagger \ket{\psi_\t{i}} + h.c. \right].
\end{align*}
Note that both rotating-wave terms 
(first line) and counter-rotating-wave terms
(second line)
contribute in general, weighted by $I_\mp$ respectively.

\begin{figure}[t!]
	\centering
	\includegraphics[width=0.8\columnwidth]{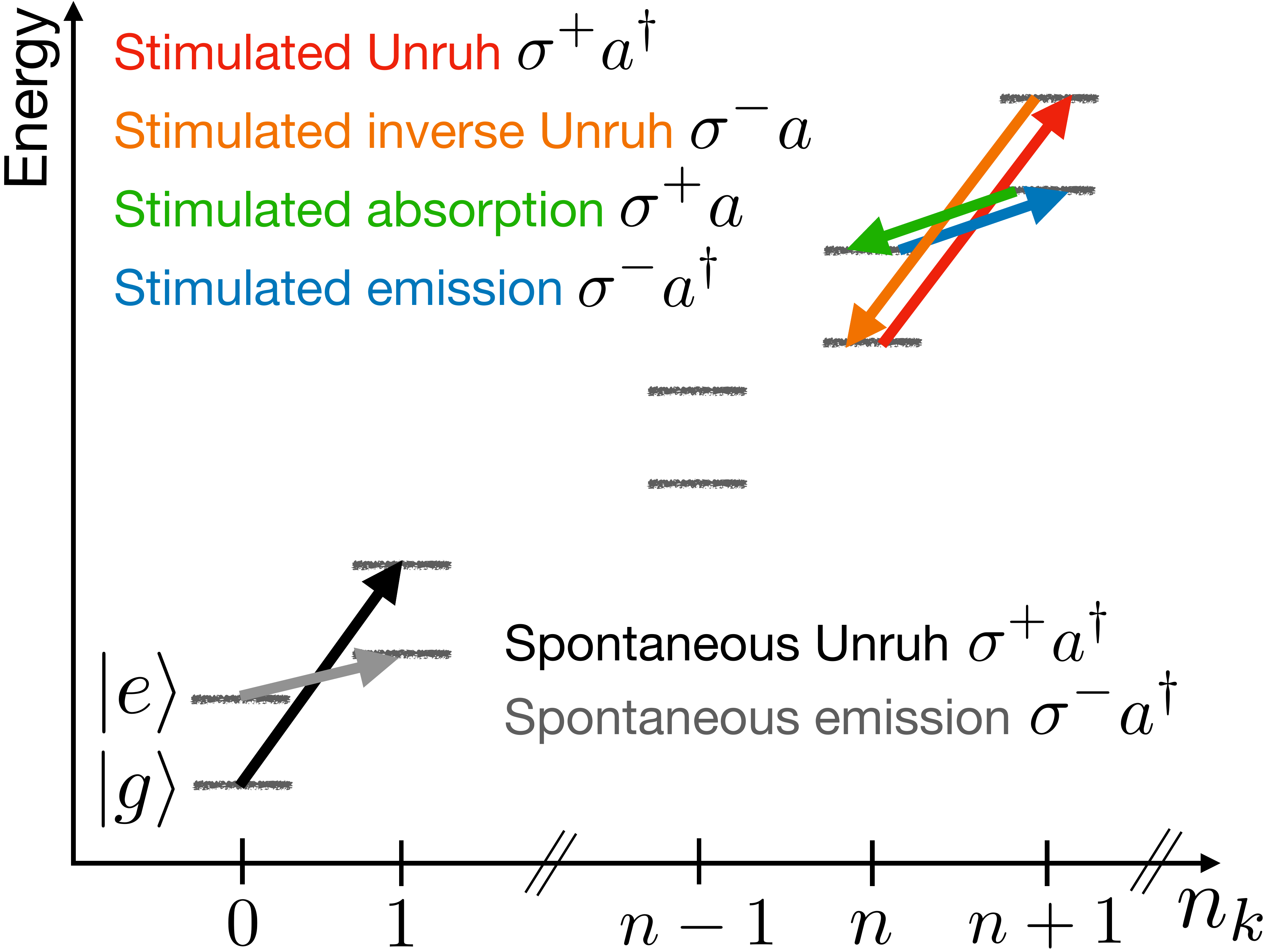}
	\caption{\label{fig1}
		Comparison between conventional first order processes in light-matter interaction
		that happens in inertial motion (gray, green, blue), and those that happen only in the presence of 
		non-inertial motion (black, orange, red). Shown here are the pairs of states 
		$\{\ket{g}\otimes \ket{n_k}, \ket{e}\otimes \ket{n_k}\}$ as the field photon number $n_k$ increases,
		and the transitions between them due to the various processes together with the terms that describe
		the processes.
	}
\end{figure}

\emph{Conventional inertial phenomena.} 
Oft-studied phenomena \cite{Dirac27} such as 
spontaneous emission (i.e., $\ket{e,0} \rightarrow \ket{g,1_\bldk}$), 
stimulated emission (i.e., $\ket{e,n_\bldk} \rightarrow \ket{g,(n+1)_{\vb{k}}}$), and 
absorption (i.e., $\ket{g,(n+1)_\bldk} \rightarrow \ket{e,n_\bldk}$) 
all arise from the rotating-wave terms when the atom is in inertial motion (see \cref{fig1}).
(Here, $\ket{n_\bldk} = (n!)^{-1/2}\hat{a}_\bldk^\dagger \ket{0}$ is the Fock state of the field.)
Indeed in inertial motion --- given by the trajectory $x^\mu (\tau) = (\gamma \tau, \gamma \vb{v}\tau)$ 
where $\gamma=(1-\abs{\vb{v}}^2)^{-1/2}$ ---the emission and absorption amplitudes,
\begin{equation}\label{inertialA}
\begin{split}
	\mathcal{A}_{\ket{e,n}\rightarrow\ket{g,n+1}}(\vb{k}) 
		&= G \sqrt{n+1}\cdot \delta[\Omega - \gamma(\omega_\bldk - \vb{k}\cdot \vb{v}) ]  \\
	\mathcal{A}_{\ket{g,n+1}\rightarrow\ket{e,n}}(\vb{k}) 
		&= G \sqrt{n}\cdot \delta[\Omega -\gamma(\omega_\bldk- \vb{k}\cdot \vb{v}) ]
\end{split}
\end{equation}
are non-zero only on resonance, i.e. when the excitation energy of the atom matches the (relativistically
Doppler-shifted) energy of the photon. 
This implies that in inertial motion, emission and absorption processes are strictly
due to the rotating-wave terms in the interaction.
Here, the $\delta$ distribution arises from the integral $I_-$, which quantifies the
contribution of the rotating-wave terms.

For later reference we note that the probabilities for these processes satisfy the Einstein relations:
\begin{equation}\label{inertialEinstein}
\begin{split}
	\abs{\mathcal{A}_{\ket{e,n}\rightarrow\ket{g,n+1}}}^2 
		&= (n+1)\times \abs{\mathcal{A}_{\ket{e,0}\rightarrow\ket{g,1}}}^2 \\
	\abs{\mathcal{A}_{\ket{g,n+1}\rightarrow\ket{e,n}} }^2 
		&= n\times \abs{\mathcal{A}_{\ket{e,0}\rightarrow\ket{g,1}} }^2
\end{split}
\end{equation}
i.e. stimulated emission and absorption are respectively $(n+1)$ and $n$ times more probable
than spontaneous emission. 

\emph{Conventional (i.e. spontaneous) Unruh effect.}
Within this context, the conventional Unruh effect \cite{Unruh76,Tak86} is concerned with the possibility
that an atom in its ground state accelerating through the field vacuum gets excited. 
Note that the excitation of a ground state atom in vacuum is impossible in inertial motion.
For a general trajectory $x^\mu(\tau)$, the amplitude for the conventional (i.e. spontaneous) Unruh process is,
\begin{equation*}
	\mathcal{A}_{\ket{g,0}\rightarrow \ket{e,1}}(\vb{k}) = G \int \dd \tau\, e^{i \Omega \tau + i k^\mu x_\mu(\tau)}  
	= G\, I_+(\Omega,\bldk).
\end{equation*}
For inertial motion, we have that, $I_+ = \delta[\Omega + c k -\vb{k}\cdot \vb{v}]$, which does not contribute
since $c > \abs{\vb{v}}$ always; so in inertial motion, the above amplitude is identically zero.
But this need not be so for non-inertial motion.
Indeed for the specific case of uniform acceleration of magnitude $a$, i.e.
$x^\mu(\tau) = (\sinh(a\tau)/a, \cosh(a\tau)/a,0,0)$, we have that \cite{UnWal84,Tak86,Ben19},
\begin{equation}\label{unurhP}
	\abs{\mathcal{A}_{\ket{g,0}\rightarrow \ket{e,1}}}^2 = G^2 \frac{2\pi/(\Omega a)}{e^{k_B T_\t{U}/\Omega}-1};
	\quad T_\t{U} = \frac{a}{2\pi k_B},  
\end{equation}
which is a Bose-Einstein distribution at temperature $T_\t{U}$. This is the crux of the conventional
Unruh effect --- an atom uniformly accelerated through the vacuum perceives an apparently thermal field.

We note a few key aspects. First, the probability of the spontaneous Unruh process 
is non-zero in a non-inertial frame precisely because of the time-dependence of the (here, exponential) Doppler shift seen by the atom 
(which is the physical interpretation of the integral $I_+$, in analogy with the corresponding integral 
in the inertial case in \cref{inertialA}).
This makes clear that even for realistic trajectories, that do not involve eternal uniform acceleration,
the Unruh process has a non-zero probability. 
Second, the energy required to simultaneously excite the atom and create a photon comes from 
the accelerating agent; indeed in a more complete treatment, the excitation of the detector is accompanied
by a recoil of the atom's center-of-mass degree of freedom \cite{Pad85,Paren95,SudKemp21}.
Third, despite it being a robust quantum feature of accelerated 
bodies \cite{Sew82}, the Unruh temperature (restoring constants), 
$T_U = \hbar a/(2\pi k_B c) \approx 10\, \t{mK} \left( \frac{a}{10^{18}\, \t{m/s^2}} \right)$, 
has so far rendered its experimental study infeasible.
(It is worth noting that the term `Unruh effect' is sometimes
reserved for the class of trajectories for which the accelerated system is driven to a thermal or near-thermal state.
We will consider more general trajectories and will call any excitation of quantum systems due to non-inertial motion
an Unruh effect.)

\emph{Stimulated Unruh effect.} Given the central importance of the conventional Unruh process, it is worth enquiring whether
more general variations on the theme exists, and whether these variations may be amenable to experimental
observation. Deriving inspiration from the stimulated processes that happen in inertial 
motion [\cref{inertialA}], we consider the possibility of stimulating the Unruh process.
That is, instead of the initial state $\ket{g,0}$, we consider the state $\ket{g,n_\bldk}$.
The interaction leads to the transformation,
\begin{equation}\label{stateTrans}
\begin{split}
	\ket{g,n_\bldk} \rightarrow 
	G \Big[ 
		& I_+\, \sqrt{n+1}  \ket{e,(n+1)_\bldk} \\
		& \quad + I_-\, \sqrt{n} \ket{e,(n-1)_\bldk}		
	\Big];
\end{split}
\end{equation}
here the transformed state is unnormalized for brevity.
The first term ($\propto I_+$) corresponds to a stimulated Unruh process, which arises from counter-rotating terms in the interaction,
is absent in inertial motion, and does not depend on the stimulating photon being resonant with the atom.
The second term ($\propto I_-$) corresponds to conventional (resonant) absorption, which is due to rotating wave terms in
the interaction, and therefore relies on atom-photon resonance. 
The stimulated Unruh process stands in the same relation to the conventional (spontaneous) Unruh process
in an accelerating frame, as conventional stimulated emission is to spontaneous emission in an inertial frame.
Importantly, the probability of the stimulated Unruh processes is enhanced by a 
factor of $n+1$ compared to the spontaneous version:
\begin{equation}
	\abs{\mathcal{A}_{\ket{g,n}\rightarrow \ket{e,n+1}}}^2 
		= (n+1)\times \abs{\mathcal{A}_{\ket{g,0}\rightarrow \ket{e,1}}}^2,
\end{equation}
where $\abs{\mathcal{A}_{\ket{g,0}\rightarrow \ket{e,1}}}^2$ is the probability for the 
spontaneous Unruh process (for uniform acceleration, given in \cref{unurhP}). 
This equation elicits the well known Einstein relation
[\cref{inertialEinstein}] for inertial stimulated emission.

The scaling with photon number immediately suggests that the experimental obstruction to observing
the spontaneous Unruh effect can be alleviated. However two aspects need to be addressed.
First, in order to take advantage of the $n-$scaling, realistic experiments would need to use a large
mean photon number, whereas large-$n$ Fock states are challenging to prepare. 
Second, the state transformation in \cref{stateTrans} produces an undesirable resonant
absorption component in addition to the stimulated Unruh process.
In the following we address both issues.

\emph{Stimulation with general field states.} We choose to represent 
general field states in terms of the (over-)complete coherent state basis \cite{Sud63,Glaub63}.
The transition amplitude for the stimulated Unruh process where the field is in 
these basis states is (up to a phase factor),
\begin{equation*}
	\mathcal{A}_{\ket{g,\alpha}\rightarrow \ket{g,\beta}}(\bldk) 
		= \frac{G e^{-\abs{\alpha - \beta}^2/2}}{(2\pi)^{3/2}\sqrt{\omega_\bldk}}
		(\alpha I_- + \beta^* I_+).
\end{equation*}
Here $\ket{\alpha},\ket{\beta}$ are field coherent states, which are not orthogonal 
(i.e. $\abs{\braket{\alpha}{\beta}}^2 =e^{-\abs{\alpha - \beta}^2}$).
The probability that the atom gets excited, irrespective of the final field state in mode $\bldk$, is,
\begin{equation}\label{Palpha}
\begin{split}
	\mathcal{P}_{\alpha}(\bldk) 
	&= \int \frac{\dd^2 \beta}{\pi} \abs{\mathcal{A}_{\ket{g,\alpha}\rightarrow \ket{g,\beta}}(\bldk)}^2\\
	&= \frac{G^2}{(2\pi)^2 \omega_\bldk} \left( \abs{\alpha}^2 \abs{I_+ + I_-}^2 + \abs{I_+}^2 \right).
\end{split} 
\end{equation}
Note that when $\alpha = 0$, i.e. the initial field state in the vacuum, this expression
reduces to, $\mathcal{P}_0(\bldk) \propto \abs{G\, I_+}^2$, consistent with the conventional, i.e.,  spontaneous Unruh effect
for a general trajectory.

For an arbitrary initial field state, $\hat{\rho}_\t{i} = \int P(\alpha) \dd^2 \alpha/\pi$,
where $P(\alpha)$ is a generalized probability distribution in the coherent state basis \cite{Sud63,Glaub63}, 
the probability that the atom is excited on an arbitrary trajectory can be shown to be,
$\mathcal{P}(\bldk) = \int P(\alpha)\mathcal{P}_\alpha (\bldk) \dd^2 \alpha/\pi$, or explicitly,
\begin{equation}\label{generalPk}
	\mathcal{P}(\bldk) 
		= \frac{G^2}{(2\pi)^3 \omega_\bldk} \left( \langle\hat{a}_\bldk^\dagger \hat{a}_\bldk \rangle 
		\abs{I_+ + I_-}^2 + \abs{I_+}^2 \right).
\end{equation}
Here we have used the relation between integrals of $P$ and expectations of normal-ordered 
operators \cite{Glaub63}. The implication is that in general --- irrespective of the input field state --- 
the probability of the stimulated Unruh process grows with the average photon number.

\emph{Acceleration-induced transparency.} Note that the atom gets 
excited from three processes [see \cref{generalPk}]:
the spontaneous Unruh effect (last term $\propto \abs{I_+}^2$) --- which is challenging to observe;
the stimulated Unruh effect ($\propto \langle\hat{a}^\dagger \hat{a}\rangle \abs{I_+}^2$) --- whose probability
can be dramatically larger;
and conventional (resonant) absorption ($\propto \langle \hat{a}^\dagger \hat{a}\rangle \abs{I_-}^2$).
Except that, as \cref{generalPk} shows, stimulation amplifies resonant and non-resonant terms equally:
the actual probability is $\propto \langle\hat{a}^\dagger \hat{a}\rangle \abs{I_+ + I_-}^2$.
Excitation due to the Unruh effect appears, therefore, to be `contaminated' by conventional resonant absorption. 

In order to make the stimulated Unruh effect observable, the question arises as to whether there are any means 
to suppress $I_-$ relative to $I_+$. 
Indeed, one should expect that it is possible to make $I_-$ smaller than $I_+$ by choosing the stimulating field 
mode to be far detuned from the atomic resonance. This is because absorption is a resonant process 
whereas the Unruh process is non-resonant. 

In fact, as we now show, more is possible in the presence of acceleration: by choosing suitably accelerated trajectories, one can, in principle, completely suppress resonant absorption. 
This allows one to achieve $I_-=0$ while also $I_+\neq 0$. This phenomenon, which may be called ``acceleration-induced transparency", arises because acceleration not only activates the (non-resonant) counter-rotating terms but it also strongly modulates the (resonant) rotating terms. 
(Acceleration-induced transparency is unrelated to the so-called electromagnetically-induced transparency \cite{EIT05} which is caused by destructive interference of 
excitation amplitudes in a three-level atom induced by resonant excitation with coherent field states).

To prove acceleration-induced transparency, let us assume that an atomic gap, $\Omega$, and the stimulating mode's wave vector, $\bldk$, are chosen. Our task is to show that there are trajectories for which $I_-=0$ and $I_+\neq 0$. 
To this end, we associate to any trajectory $x_\mu(\tau)$ the `phase function', $\alpha(\tau)=k^\mu x_\mu(\tau)$, 
so that, $I_\pm (\Omega,\bldk) = \int \dd \tau\, e^{i \Omega \tau \pm i \alpha(\tau)}$.
A physical trajectory is one that is timelike, i.e. $\dot{x}_0 > 0$.
Note that the phase function of a physical trajectory satisfies: $\dot{\alpha}(\tau) > 0\, \forall \tau$.
This is because $\dot{\alpha}(\tau)$ is scalar, and in an instantaneous rest frame it reads, 
$\dot{\alpha}(\tau)= k^\mu \dot{x}_\mu(\tau)=k^0 \dot{x}_0(\tau) >0$. 
Conversely, and more importantly, to any phase function $\alpha(\tau)$ obeying $\dot{\alpha}(\tau)>0\,\forall\, \tau$, 
there exists a corresponding physical trajectory. 
We prove this by construction:
choose a coordinate system such that $k=(k_0,k_0,0,0)$; then the trajectory,
\begin{equation}
\label{phase_function_trajectory}
    \dot{x}_{\mu}(\tau)=\lb \frac{1}{2}\lb\frac{k_0}{\dot{\alpha}\lb\tau\rb}+\frac{\dot{\alpha}\lb\tau\rb}{k_0}\rb,\frac{1}{2}\lb\frac{k_0}{\dot{\alpha}\lb\tau\rb}-\frac{\dot{\alpha}\lb\tau\rb}{k_0}\rb,0,0\rb
\end{equation}
is timelike with $\tau$ its proper time (i.e.,  $\dot{x}^{\mu}\dot{x}_{\mu}=1$). It is straightforward to verify that this trajectory obeys $k^\mu \dot{x}_\mu(\tau) = \dot{\alpha}(\tau)$, i.e., that it produces the desired phase function $\alpha(\tau)$ up to an irrelevant integration constant that  translates the trajectory. 
Notice that the trajectory is inertial if $\dot{\alpha}$ is constant, and accelerating otherwise.  

Our remaining task is to find examples of phase functions $\alpha(\tau)$ obeying $\dot{\alpha}(\tau)>0~\forall \tau$ for which $I_-=0$ and $I_+\neq 0$. Through \cref{phase_function_trajectory} we then obtain corresponding trajectories for which the Unruh effect can be arbitrarily strongly stimulated while conventional absorption vanishes. 
To this end, consider a phase function satisfying,
\begin{equation}\label{pf}
	\dot{\alpha}(\tau) := k^0
	\begin{cases}
		s_0  & \tau < 0 \\
		s_0+\frac{s_1-s_0}{T_1}\tau & \tau \in [0,T_1) \\
		s_1+\frac{s_2-s_1}{T_2-T_1}\lb\tau-T_1\rb  & \tau \in [T_1, T_2) \\
		s_2 & \tau \geq T_2,
	\end{cases}
\end{equation}
where the constants $\{s_i,T_i\}$ can be chosen arbitrarily except that we require
$\dot{\alpha}>0$ and $0<T_1<T_2$. 
The corresponding trajectories are initially inertial, then possess two phases of acceleration, followed 
by inertial motion. 

\begin{figure}[t!]
	\centering
	\includegraphics[width=\columnwidth]{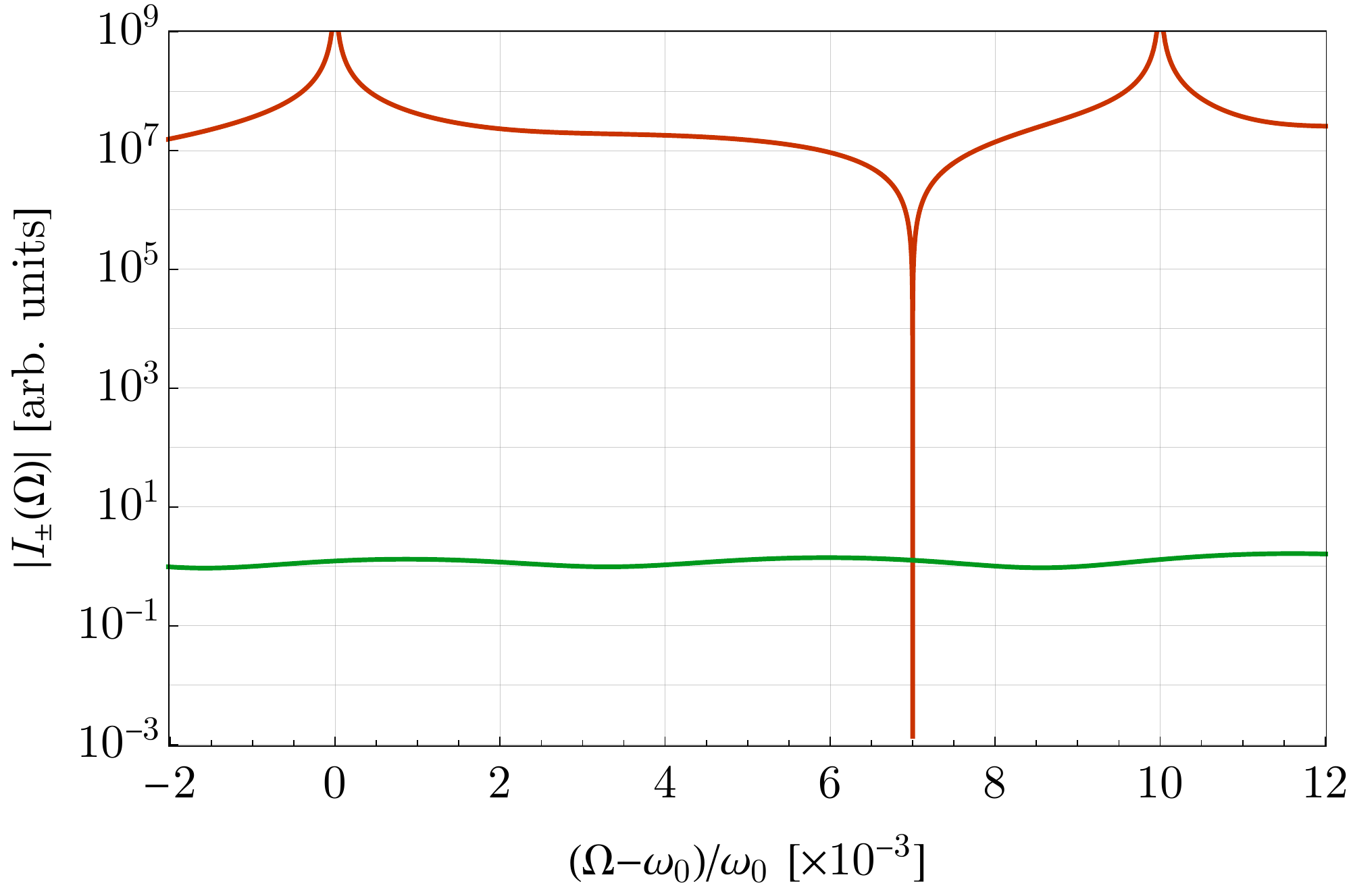}
	\caption{\label{fig2}
		The two curves display $\abs{I_\pm(\Omega)}$ for an example of a trajectory of the 
		form given in \cref{pf}. The red curve is the resonant contribution $\abs{I_-}$ which shows 
		the strength of conventional absorption while the green curve is the
		non-resonant contribution $\abs{I_+}$ which shows the strength of Unruh-type counter-rotating effects. 
		Importantly, we see that the resonant contribution dips below that of the latter 
		(here $\abs{I_-/I_+}\approx 10^{-3}$), so that at that frequency, the probability of resonant absorption
		is $\sim 10^{-6}$ of the stimulated Unruh process. 
		The two peaks in $I_-$ correspond to absorption at the Doppler shifted frequencies 
		$\omega_{0} = k^0 s_{0}$, $\omega_{2} = k^0 s_{2}$ that are due to the initial and final inertial 
		velocities of the trajectory.
	}
\end{figure}

The question is whether among these trajectories there are ones which exhibit acceleration-induced transparency 
for some gap $\Omega$. To produce explicit examples, we held the parameters $s_0,s_2$ and $T_1$ fixed and 
varied the parameters $s_1$ and $T_2$. In the $(s_1,T_2)$-plane, we plotted the curves for which 
$\t{Re}\left[I_-\right]=0$ or $\t{Im}\left[I_-\right]=0$. These curves intersect, which shows that there 
are parameter values for which the corresponding trajectory possesses acceleration-induced transparency at 
the chosen gap $\Omega$. This means that it is possible, in principle, to make the stimulated Unruh process 
dominate arbitrarily strongly over all conventional, i.e., resonant processes, at this order in perturbation 
theory and for a given gap $\Omega$.

In \cref{fig2}, we plot $\abs{I_\pm (\Omega)}$ for such a trajectory. 
The plot shows that the resonant effects described by $I_-$ tend to dominate over the non-resonant effects described 
by $I_+$, except for the arbitrarily strong acceleration-induced suppression of $I_-$ at a particular 
value of $\Omega$.  Notice that this plot in effect also shows the $k_0$ dependence. 

\emph{At high intensity stimulation, the effect is catalysed.} 
The ideas described above --- the stimulated Unruh effect and acceleration-induced transparency --- 
make it possible to amplify the Unruh effect by stimulating it with large photon numbers, while
suppressing resonant absorption. When used in tandem, and when the stimulation is done by 
a field coherent state, it turns out that the resulting physical effect proceeds with little 
change to the field state. 
To see this, note that,
\begin{equation*}
	\mathcal{A}_{\ket{g,\alpha} \rightarrow \ket{e,\alpha}} (\bldk)
	\propto G \left( \alpha^* I_+ + \alpha I_-  \right)  \neq 0.
\end{equation*}
That is, the field coherent state catalyses the excitation of the atom, without itself changing.
The ratio of the probability for the catalysis of the stimulated Unruh process and the total probability for the stimulated Unruh process (when the 
field starts in a coherent state) [\cref{Palpha}] is,
\begin{equation*}
	\frac
	{\abs{\alpha^* I_+ + \alpha I_-}^2}
	{\abs{\alpha}^2 \abs{I_+ + I_-}^2 + \abs{I_+}^2}.	
\end{equation*}
Further, when the trajectory is  chosen to be one that suppresses absorption (i.e. $\abs{I_-} \ll \abs{I_+}$), 
the fraction of the catalysed process dominates the stimulated Unruh process:
\begin{equation*}
	\frac{\abs{\alpha}^2}{\abs{\alpha}^2 + 1}\left[ 1 + \mathcal{O}\left(\abs{\tfrac{I_-}{I_+}}\right) \right] 
	\xrightarrow{\abs{\alpha}\gg 1} 1 + \mathcal{O}\left(\abs{\tfrac{I_-}{I_+}}\right).
\end{equation*}
Physically, this is to be expected since, in this scenario, the energy needed to excite the atom is  
negligible compared to the total energy present in the photon field.

In addition to the effects considered above where the accelerating atom starts in the ground state, 
it is possible to consider the case where the atom starts in the excited state. Variants of the
stimulated Unruh effect can then be considered with this initial condition for the atom's internal state.
Such processes are time-reversed versions of the ones already studied above. Indeed, this means that
the amplitudes for such processes are obtained by replacing $I_\pm \rightarrow I_\mp^*$ in the equations
above.

\emph{Conclusions.} 
We have shown that the effects of counter-rotating, i.e., non-resonant terms can be stimulated by the presence of photons in 
the field, leading to what can be described as a stimulated Unruh effect. 
We also showed that acceleration can be used to modulate the conventionally dominant resonant, i.e., rotating wave terms. We found that by judicious choices of the accelerating trajectories, resonant absorption can even be switched off. 
In this fashion, the effect of counter-rotating wave terms --- conventionally neglected in the description
of the weak-coupling regime --- can be made to be the dominant (and, in principle, even the \emph{sole}) mode of light-matter interaction.
This shows that, with strong stimulation, the acceleration of an atom can dramatically affect
the response of its internal state. This opens new avenues to measuring the acceleration-induced activation of the counter-rotating terms of the light-matter interaction even in the weak-coupling regime, for example
by measuring the recoil of the atom \cite{SudKemp21}, or by directly measuring the internal state under strong stimulation.  Detailed calculations for realistic scenarios with the electromagnetic light-matter interaction, instead of the idealized Klein Gordon interaction 
are in progress.  

It will also be very interesting to investigate analogs of our results that are gravity-induced rather than acceleration-induced.  

\emph{Acknowledgements.}
AK acknowledges support through the Discovery Grant Program of the National Science and Engineering Research Council of Canada (NSERC), the Discovery Grant Program of the Australian Research Council (ARC) and a Google Faculty Research Award. Research at Perimeter Institute is supported in part by the Government of Canada through the Department of Innovation, Science and Industry Canada and by the Province of Ontario through the Ministry of Colleges and Universities. 

\bibliographystyle{apsrev4-1}
\bibliography{refs_Unruh}

\end{document}